\documentclass[a4paper,11pt]{article}
\usepackage{graphicx}
\usepackage{dcolumn}
\usepackage{bm}
\usepackage{epsfig,amsmath}
\usepackage{amssymb}
\usepackage{subfigure}
\usepackage[normalem]{ulem}
\usepackage[usenames,dvipsnames]{color}
\usepackage[pagebackref=false, colorlinks=true]{hyperref}
\definecolor{redish}{rgb}{0.7,0.2,0.0}  
\definecolor{bluish}{rgb}{0.2,0.5,0.8}

\hypersetup{linkcolor=redish,          
                  citecolor=blue,        
                  filecolor=magenta,      
                  urlcolor=bluish}          

\DeclareFontFamily{U}{rsfs}{}         
\DeclareFontShape{U}{rsfs}{m}{n}{<5> rsfs5 <6><7> rsfs7          %
  <8><9><10><10.95><12><14.4><17.28><20.74><24.88> rsfs10}{}     %
\DeclareMathAlphabet{\mathfs}{U}{rsfs}{m}{n}

\def \f{\frac}

\def \o{\omega}

\def \a{\alpha}

\def \s{\sigma}

\def \T{\Theta}

\title{Near- and sub-solar-mass naked singularities and black holes from transmutation of white dwarfs}

\author{Chandrachur Chakraborty$^1$\footnote{chandrachur.c@manipal.edu}    ~ ~and~~ Sudip Bhattacharyya$^2$\footnote{sudip@tifr.res.in} 
 \\ 
 {$^1$\it Manipal Centre for Natural Sciences, Manipal Academy of Higher Education, Manipal 576104, India }
 \\
 $^2$\it Department of Astronomy and Astrophysics,\\
\it Tata Institute of Fundamental Research, Mumbai 400005, India}

\date{}

\begin{document}

\maketitle

\begin{abstract}
Recent gravitational wave events have suggested the existence of near-solar-mass black holes which cannot be formed via stellar evolution.
This has opened up a tantalizing possibility of future detections of both black holes and naked singularities in this mass range. Existence of naked singularities is a topical and fundamental physics issue, but their formation mechanism is not yet clear. Here, we show that some white dwarfs can realistically transmute into black holes and naked singularities with a wide range of near- and sub-solar-mass values by capturing 
asymmetric or non-self-annihilating primordial dark matter (PDM) particles. We argue that, while a type Ia supernova due to the accumulation of dark matter at the core of a white dwarf could also be a possibility, the transmutation of a white dwarf into a black hole or a naked singularity is a viable consequence of the capture of non-self-annihilating PDM particles.
These white dwarf transmutations can have a significant role in probing the physics of dark matter and compact objects, and could be tested using the rates and locations of mergers over the cosmological time scale.
\end{abstract}

\section{Introduction}
The gravitational wave events GW190425 \cite{gw1} and GW190814 \cite{gw2} could have involved mergers of near-solar-mass  black holes (BHs). This has recently created an interest in such BHs, which cannot be formed via stellar evolution \cite{dasg}.
While a popular candidate of such low-mass BHs are so far observationally not confirmed primordial
black holes (PBHs; \cite{dasg}), an alternative process to form low-mass BHs, involving the capture of primordial dark matter (PDM) particles by a cosmic host and the subsequent transmutation of the host, has been proposed \cite{dasg}.
GW190425 and GW190814 opened up the possibility of reliable detections of not only solar-mass BHs but also naked singularities and sub-solar mass black holes with advanced Laser Interferometer Gravitational-Wave Observatory (aLIGO;  \cite{aligo}) and Laser Interferometer Space Antenna (LISA) \cite{lisa} in the future.
In this paper, we show that the transmutation of neutron stars (NSs), which is widely discussed, may not give rise to naked singularities and sub-solar mass BHs, but a whole range of sub-solar and slightly
super-solar mass Kerr BHs and Kerr naked singularities could be formed from the transmutation of some white dwarfs (WDs).

An astrophysical collapsed object is characterized by two parameters: mass $M$ and angular momentum $J$. When the dimensionless Kerr parameter $(a)$ satisfies\footnote{$G$ is the Newton's gravitational constant and $c$ is the speed of light in the vacuum.} $a=Jc/(GM^2) \leq 1$ \cite{kerr, cb17, bcb19}, the Kerr singularity remains hidden inside the event horizon and is known as a Kerr black hole (BH). On the other hand, the event horizon does not exist if $a > 1$, and the object is called a Kerr naked-singularity (or superspinar) which could be visible to a distant observer \cite{ckj}. 
Note that the cosmic censorship hypothesis, conceived by Penrose \cite{penrose}, states that a gravitational singularity would remain hidden inside an event horizon. But, later theoretical works found the tantalizing possibility of existence of naked singularities (see \cite{psj, gh, jmn, ckj, ss, jm, gss, jsy, wag} and references therein).

The existence of such objects remains an open, topical and fundamental issue (see \cite{ckj, ckp} and references therein), which has recently been discussed for SgrA* \cite{eht2022, vag}, M87* \cite{bambi, gcyl} and GRO J1655-40 \cite{cbgm, cbgm2}. 
Although some possible formation mechanisms of naked singularities were proposed earlier  \cite{js, lb, nakao}, it is still unclear, in which realistic situation a naked-singularity could form. 
Particularly, to the best of our knowledge, no mechanism for the formation of a near-solar-mass naked-singularity has been proposed.

Our proposed new channel to form such naked-singularities assumes the existence of asymmetric or non-self-annihilating PDM particles \cite{ste}. There are several proposed DM candidates and 
non-self-annihilating PDM particles are one of them (see Fig. 1 of \cite{gel} for details).
It is suggested that the interaction between PDM particles and regular baryonic matter is extremely small except gravity. 
Thus, non-self-annihilating PDM particles could enter a stellar object/host due to gravity, lose energy via scattering, attain thermal equilibrium with the stellar constituents to join an isothermal sphere around the core (known as thermalization \cite{dasg20}), become bound to the host, and accumulate at the core of the host \cite{dasg}.
If a sufficiently large number of PDM particles are captured, a tiny BH should form at the core of the host, for example a neutron star (NS) \cite{dasg}. After formation, this endoparasitic BH immediately starts accreting matter from its host \cite{dasg}, and after a certain time the entire host could transmute into a collapsed object, i.e., either a BH or a naked singularity.
Not only NSs, but other cosmic objects (e.g., WDs, stars, planets,  etc.) could also  capture non-self-annihilating PDM particles continuously, and could transmute into collapsed objects
by the same process.
A necessary condition for the host for having already been transmuted into a collapsed object is 
$t_{\rm d} < {\rm Min}(t_{\rm U}, t_{\rm LT})$, which depends on various parameter values.
Here, $t_{\rm d}$ is 
the transmutation time, $t_{\rm U}$ ($\approx 1.38 \times 10^{10}$ yr \cite{pl3}) is the current age of the Universe, and $t_{\rm LT}$ is the maximum possible lifetime of the host.
Whether the transmuted collapsed object will be a Kerr BH or a Kerr naked singularity depends on the initial host parameter values, and how much mass and angular momentum
leave the system during accretion by the endoparasitic BH from its host.

\section{Methodology}~
A general methodology of transmutation for all stellar and planetary hosts is discussed here in detail. 

\subsection{\label{s1}A condition to transmute into a naked singularity}

A necessary condition for transmutation of a host into a naked singularity is
$a_{\rm h} = J_{\rm h}c/(GM_{\rm h}^2) > 1$, where $a_{\rm h}$, $J_{\rm h}$, and $M_{\rm h}$ are the dimensionless Kerr parameter or spin parameter, angular momentum, and mass of the host, respectively.
If the moment of inertia of the host is $I_{\rm h}=AM_{\rm h}R_{\rm h}^2$ ($A$ is a constant), one can write from $J =2\pi I_{\rm h}/T_{\rm h}$ ($T_{\rm h}$ is the host spin period):
\begin{eqnarray}
 a_{\rm h}=\f{2\pi AcR_{\rm h}^2}{GM_{\rm h}T_{\rm h}}.
 \label{feq}
\end{eqnarray}
Therefore, a necessary condition to form a naked-singularity is
\begin{eqnarray}
 T_{\rm h} < \f{2\pi A cR_{\rm h}^2}{GM_{\rm h}}.
 \label{feq1}
\end{eqnarray}

\subsection{\label{s2}Calculation of the transmutation time}

Here, we discuss the calculation of the transmutation time $t_{\rm d}$.
Let us consider that a host begins to capture the non-self-annihilating PDM particles at time $t=0$, and a tiny BH of mass $M_0$ is formed at time $t=t_0$ in the core of the host. The capture rate $F$ \cite{ktwd, mcd} of the PDM particles depends linearly with the DM density $\rho$ \cite{ktwd, mcd} and inversely with the velocity dispersion $\bar{v}(=220$ km/s). 
With fixed values of mass and radius of a host, $F$ also depends on the mass ($m_{\chi}$) of a PDM particle and PDM particle-nucleon scattering cross section ($\s$). 
$F$ does not depend on $\s$ \cite{bell} for $\s > \s_{\rm cr} \sim mR_{\rm h}^2/M_{\rm h}$, where $\s_{\rm cr}$ is the critical cross section and $m$ is the mass of one baryon of the host. 
A tiny BH is formed when the PDM particles satisfy the collapse criterion: $N \geq {\rm Max}[N_{\rm Ch}, N_{\rm self}]$ \cite{ktwd, mcd, dasg} where $N$ is the total number of  PDM particles accumulated inside the host. $N_{\rm Ch}$ and $N_{\rm self}$ are the Chandrasekhar limit and the number of PDM particles required for initiating the self-gravitating collapse, respectively \cite{ktwd}. If the PDM density is larger than the baryon density within the thermal radius of a host, the PDM particles become self-gravitating. 
Thus, one needs to estimate the thermalization timescale $t_{\rm th}$ \cite{ace, mcd} which could play a crucial role for calculating the formation timescale. In fact, the PDM particles have to attain the thermal equilibrium with the stellar constituents to accumulate at the core of the host.

With the above information, one can calculate the formation 
timescale  $t_0^{\rm b}=t_{\rm th}+{\rm Max}[N_{\rm Ch}^{\rm b}, N_{\rm self}]/F$ (or, $t_0^{\rm f}=t_{\rm th}+{\rm Max}[N_{\rm Ch}^{\rm f}, N_{\rm self}]/F$) of an endoparasitic BH due to the accumulation of bosonic (or, fermionic) PDM particles (b: boson, f: fermion) \cite{mcd, ktwd} where $N_{\rm Ch}^{\rm b} = 1.5 \times 10^{34} (100~{\rm GeV}/m_{\chi})^2$ and $N_{\rm Ch}^{\rm f} = 1.8 \times 10^{51} (100~ {\rm GeV}/m_{\chi})^3$ \cite{mcd}.

After the endoparasitic BH is formed at $t=t_0$, it begins to accrete matter from its host.
The accretion timescale $t_{\rm acc}$ is calculated as $c_{\rm s}^3 R_{\rm h}^3/(3 G^2 M_{\rm h} M_0)$ \cite{geno, mcd} ($c_{\rm s}$ is the speed of sound\footnote{The values of $c_{\rm s}$ at the core of different stellar objects could be found in \cite{balb}.} in the core of a host) assuming the spherical Bondi accretion, where $M_0=m_{\chi}N$. 
Note that, although the Bondi accretion is not exactly valid (as the matter of a host possesses a non-zero angular momentum), this assumption has no significant effect in estimation of the accretion or destruction timescale \cite{kt, rbs} even for the rapidly spinning stars \cite{el}. However, taking into account the non-self-annihilating PDM particles accumulation timescale, thermalization timescale and accretion timescale, here we provide a clear picture of the transmutation timescale $t_{\rm d}=t_0+t_{\rm acc}$ comparing to the previous papers \cite{mcd, ktwd, ace}.

\section{Results: application to white dwarfs}~
Here, we apply the above formulation only to WDs as we will show the following: (i) some WDs can transmute into collapsed objects with $t_{\rm d} < {\rm Min}(t_{\rm U}, t_{\rm LT})$; and (ii) unlike NSs, which could give rise to only somewhat super-solar mass BHs by transmutation, WDs could transmute into BHs and naked singularities with a range of sub-solar and somewhat super-solar masses. 
We consider that the angular momentum and mass, and hence
the Kerr parameter of the WD, do not change. However, we also mention the possibility of angular momentum
and mass loss (due to accretion by an endoparasitic BH), and its implications. In case there is increase of angular momentum due to accretion from a companion
star, the Kerr parameter of the WD should increase and hence it could more likely become a naked singularity.

A WD could transmute into a Kerr naked singularity, if its initial Kerr parameter ($a_{\rm WD}$) value is greater than 1, and $a_{\rm WD}$ does not decrease to below 1 during the accretion by the endoparasitic BH.
This could happen if the mass ejection from the system during the accretion is relatively small \cite{el}.

The condition $a_{\rm WD} > 1$ corresponds to 
$T_{\rm WD} < 2\pi A cR_{\rm WD}^2/(GM_{\rm WD})$ (see Eq. \ref{feq1}).
Here, $M_{\rm WD}$, $R_{\rm WD}$, and $T_{\rm WD}$ are mass, radius and spin period of the WD, respectively, and $A \sim 0.4$ \cite{bha17}.
As $R_{\rm WD}$ could be expressed in terms of $M_{\rm WD}$ using the following expression \cite{pr}:
\begin{eqnarray}
 R_{\rm WD} \approx 0.0126 R_{\odot} \left(\f{M_{\odot}}{M_{\rm WD}}\right)^{1/3}\left[1- \left(\f{M_{\rm WD}}{M_{\rm Ch}}\right)^{4/3} \right]^{1/2},
 \label{rwd}
\end{eqnarray}
(where, $M_{\rm Ch}=1.456M_{\odot}$ is the Chandrasekhar limit),
one can express $a_{\rm WD} > 1$ as $T_{\rm WD} < f(M_{\rm WD})$, where $f$ is a function of $M_{\rm WD}$.
Therefore, as argued earlier, $T_{\rm WD} < f(M_{\rm WD})$ is a necessary condition for a WD to transmute into a naked-singularity.
The observed $M_{\rm WD}$ and $R_{\rm WD}$ ranges are $\sim 0.17M_{\odot}$  \cite{her} to $1.35M_{\odot}$ \cite{cai}, and $\sim 0.002R_{\odot}$ \cite{mer} to $0.265R_{\odot}$ \cite{sok}, respectively, where $M_{\odot}$ and $R_{\odot}$ are solar mass and radius, respectively.
However, the majority of $M_{\rm WD}$ values lie between $0.5M_{\odot}-0.7M_{\odot}$ \cite{kep}. 
The measured $T_{\rm WD}$ distribution for the fast-spinning WDs peaks at $2-3$ hr \cite{kilic}, and some of them are as short as 
$\lesssim 5$ minute \cite{kilic, mont, daf}. 

\begin{figure}
 \begin{center}
{\includegraphics[width=3in,angle=0]{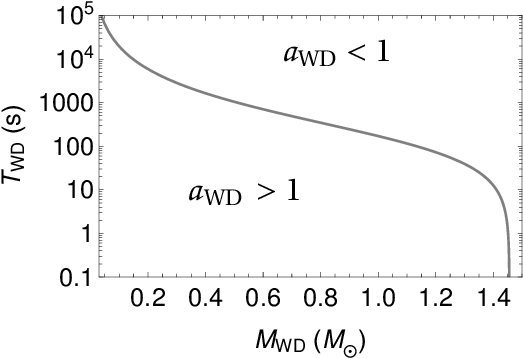}}
\caption{\label{fwda}
White dwarf (WD) spin period ($T_{\rm WD}$) versus mass ($M_{\rm WD}$) plot.
The curve represents the Kerr parameter $a_{\rm WD}=1$ for WDs. 
The parameter space above the curve is for $a_{\rm WD}<1$, while that below the curve is for $a_{\rm WD}>1$.
Thus, for the transmutation of a WD into a collapsed object (black hole ($a_{\rm WD} \leq 1$) or naked singularity ($a_{\rm WD}>1$)), the collapsed object could be a naked singularity for a large WD parameter space (see text). To form a naked-singularity of one solar mass, a solar-mass WD of $T_{\rm WD} \lesssim 170$ s is required.}
\end{center}
\end{figure}

The above mentioned ranges and Fig.~\ref{fwda} indicate what realistic values of $T_{\rm WD}$ and $M_{\rm WD}$ could lead to Kerr naked-singularities. 
For example, a fast spinning WD of $T_{\rm WD} \sim 5$ minute with a typical $M_{\rm WD} \sim 0.7M_{\odot}$ and $R_{\rm WD}\sim 0.01R_{\odot}$ could transmute into a naked-singularity of $a \approx 1.6$. 
The fastest-spinning WD LAMOST J024048.51+195226.9 with $T_{\rm WD} \approx 25$ s has been discovered recently \cite{pel}. Such a source (with $M_{\rm WD} \sim 0.7M_{\odot}$ and $R_{\rm WD} \sim 0.01R_{\odot}$) could transmute into a Kerr naked-singularity of $a \approx 19$.
Note that according to Fig.~\ref{fwda} and considering usual $T_{\rm WD}$ values, naked singularities could form typically from some of the sub-solar WDs. 

 For the strongly magnetized WDs, a new super-Chandrasekhar mass limit ($M_{\rm superCh}=2.58M_{\odot}$ \cite{dm}) was proposed, which might explain the recent observations of several peculiar Type Ia supernovae \cite{dm}. Many of the super-Chandrasekhar WDs could be fast-spinning \cite{kalita}. Thus, for instance, a WD with $M_{\rm WD}=1.639M_{\odot}$ (or, $1.96M_{\odot}$), $R_{\rm WD} \sim 987.9$ km (or, $8443.6$ km) and $T_{\rm WD} \sim 1.4$ s (or, $10.67$ s) \cite{kalita} could transmute into a naked-singularity of $a \sim 2.41$ (or, $19.32$). 
However, these  examples of strongly magnetized WDs are for the purpose of demonstration only {\footnote{But, \cite{cham} have shown that the magnetized super-Chandrasekhar white dwarfs are unlikely to be stable.}}.
Our result is applicable to all WDs irrespective of magnetic field values, and the uncertainties in the WD parameter values do not change our main conclusions.

The transmuted collapsed object from a WD would be a BH, if the final Kerr parameter value is less than 1. This could happen for initial $a_{\rm WD} < 1$ and/or sufficient mass ejection from the system during the accretion by the endoparasitic BH.
Depending on the initial $M_{\rm WD}$ value and the amount of mass ejection, the final BHs formed by this process could have a range of sub-solar and slightly super-solar masses (see Fig.~\ref{fwda}).

However, a WD could actually transmute into a collapsed object if $t_{\rm d} < {\rm Min}(t_{\rm U}, t_{\rm LT})$, where the  transmutation timescale $t_{\rm d}$ is determined by the endoparasitic BH formation timescale (including thermalization timescale) and the accretion timescale. For a WD, $t_{\rm th}$ is estimated by adding the first and second thermalization timescales, i.e., Eqs. (14) and (25) of \cite{ace}. We assume that a PDM particle is captured after a single scattering, and, hence, the PDM particles with a mass ($m_{\chi}$) range from $\sim 10$ GeV to a few $100$ TeV are considered
\cite{dasgwd, dasg20} here for the purpose of demonstration. In case of non-self-annihilating bosonic PDM particles, the BH formation in the core of a WD is determined by the number $(N_{\rm self})$ of PDM particles required for initiating a self-gravitating collapse, i.e.,
\begin{eqnarray}
 N_{\rm self}=5.6 \times 10^{47} \left(\f{100~{\rm GeV}}{m_{\chi}}\right)^{5/2}\left(\f{\T}{10^6~{\rm K}}\right)^{3/2}
 \label{nwd}
\end{eqnarray}
(the core temperature $\T$ is taken as $\sim 10^6$ K \cite{mub} for a WD), whereas for non-self-annihilating fermionic PDM particles, the BH formation is determined by the Chandrasekhar limit ($N^{\rm f}_{\rm Ch}$) for fermions.
Note that the total captured mass of PDM particles is $\sim 10^{24}-10^{17}$ kg at $t=t_{\rm U}$ with the PDM particle-nucleon scattering cross section $\s > 10^{-39}$ cm$^2$ for the above-mentioned $m_\chi$ range.
This is negligible compared to $M_{\rm WD}$ and hence should not significantly affect $a_{\rm WD}$. 
In this study, we discuss our results by considering three different DM density profiles. Two of them are most commonly used DM density profiles: Navarro-Frenk-White (NFW) and Einasto, which are expressed as follow \cite{jet, nfw},
\begin{eqnarray}
 \rho (r) &=& \rho_0 ~(r/r_s)^{-\a}(1+r/r_s)^{-3+\a} \hspace{1cm}  {\rm (NFW)}
 \label{nfw}
 \\
 \rho (r) &=& \rho_0 ~{\rm exp}\left[-2/\a~ ((r/r_s)^{\a}-1)\right] \hspace{1cm} {\rm (Einasto)}
 \label{ein}
\end{eqnarray}
where, $r$ is the distance from the Galactic center. We consider the local DM density $\rho_0=0.4$ GeV.cm$^{-3}$ following \cite{nfw, jet}, the scale radius $r_s=20$ kpc following \cite{bly} with $\a=1$ \cite{bly} (for NFW) and $\a=0.17$ \cite{jet} (for Einasto) for the purpose of demonstration. The third one is a cored DM density
profile which is expressed in the galactic plane as \cite{oll}:
\begin{eqnarray}
 \rho (r) &=& \rho_{0\rm c}/[1+(r/r_{\rm c})^2]
 \label{core}
\end{eqnarray}
where, $r_{\rm c}$ is the core radius and $\rho_{0\rm c}$ is the central density of the dark halo. We consider two different values of Table 2 of \cite{oll} for the purpose of demonstration. First one is: $\rho_{0\rm c}=0.421 M_{\odot}$ pc$^{-3} \sim 17$ GeV.cm$^{-3}$ for $r_{\rm c}=1$ kpc \cite{oll}, and another one is: $\rho_{0\rm c} =9.884 M_{\odot}$ pc$^{-3} \sim 410$ GeV.cm$^{-3}$ for $r_{\rm c}=0.2$ kpc \cite{oll}.

\begin{figure}
 \begin{center}
 \subfigure[NFW DM profile (eq. \ref{nfw}): $\rho=7 \times 10^4$ GeV.cm$^{-3}$.]{\includegraphics[width=2.8in,angle=0]
{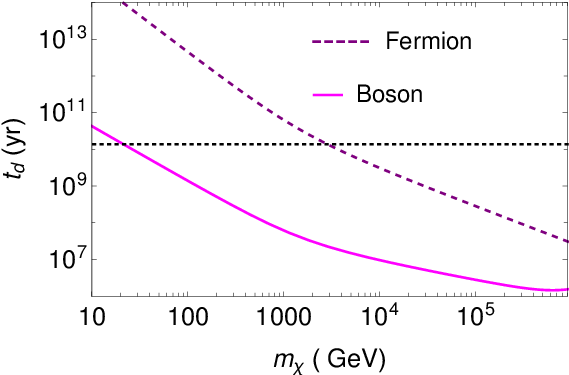}}
\hspace{0.05\textwidth}
\subfigure[Einasto DM profile (eq. \ref{ein}): $\rho=1 \times 10^4$ GeV.cm$^{-3}$.]{\includegraphics[width=2.8in,angle=0]{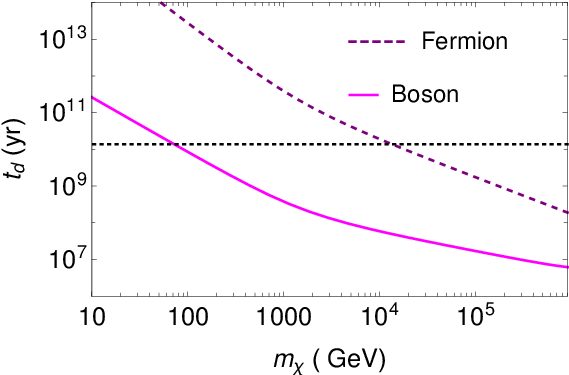}}
\subfigure[Cored DM profile (eq. \ref{core}): $\rho = 17$ GeV.cm$^{-3}$.]{\includegraphics[width=2.8in,angle=0]
{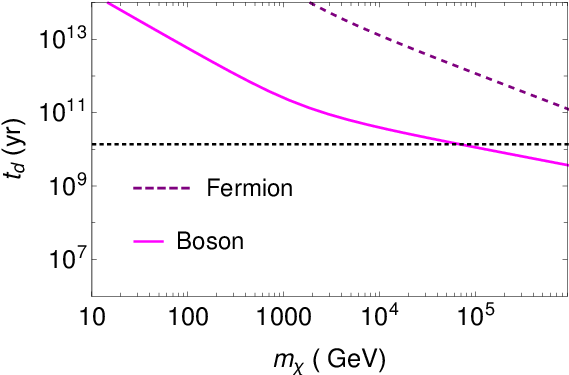}}
\hspace{0.05\textwidth}
\subfigure[Cored DM profile (eq. \ref{core}): $\rho =410$ GeV.cm$^{-3}$.]{\includegraphics[width=2.8in,angle=0]{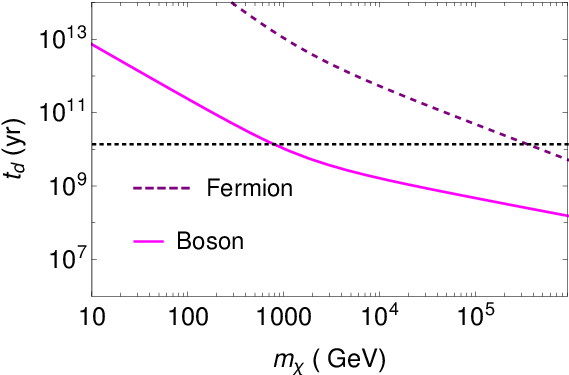}}
\caption{\label{f5}Transmutation timescale ($t_{\rm d}$) versus non-self-annihilating PDM particle mass ($m_{\chi}$) plot for the capture of different types of PDM particles and different  DM density profiles by a white dwarf (WD) of mass $0.7M_{\odot}$ and radius $0.01R_{\odot}$ with the PDM particle-nucleon scattering cross section $\s > 10^{-39}$ cm$^2$. Panels (a)-(d) show how our results vary for different DM density profiles. The black dotted horizontal line is for $t_{\rm U}$. The figure shows that there is ample possibility of transmutation of WDs into a collapsed object in a reasonable time for a wide range of $m_{\chi}$ values, depending on the value of $\rho$.
For $T_{\rm WD}=5$ minute, this collapsed object should be a Kerr naked-singularity of $a \approx 1.6$.}
\end{center}
\end{figure}

\begin{figure}
 \begin{center}
 
\subfigure[NFW DM profile (eq. \ref{nfw}): $\rho=7 \times 10^4$ GeV.cm$^{-3}$.]{\includegraphics[width=2.8in,angle=0]
{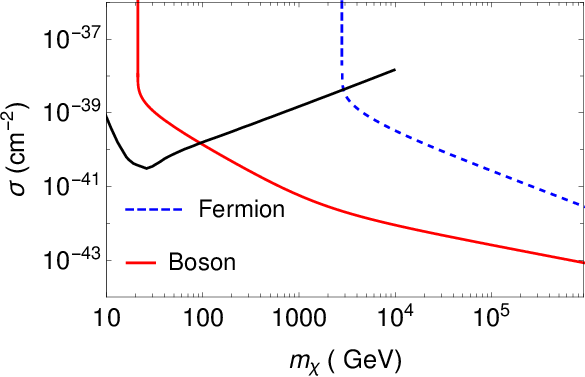}}
\hspace{0.05\textwidth}
\subfigure[Einasto DM profile (eq. \ref{ein}): $\rho=1 \times 10^4$ GeV.cm$^{-3}$.]{\includegraphics[width=2.8in,angle=0]{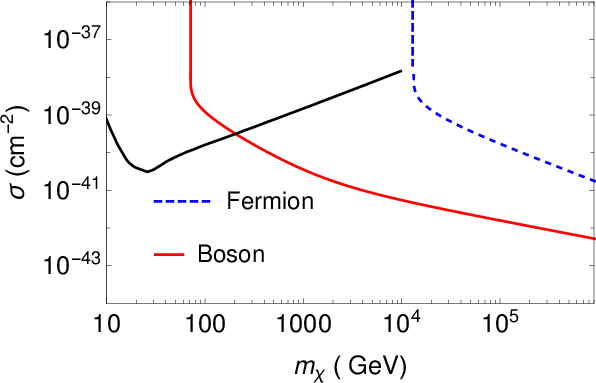}}
\subfigure[Cored DM profile (eq. \ref{core}): $\rho = 17$ GeV.cm$^{-3}$.]{\includegraphics[width=2.8in,angle=0]
{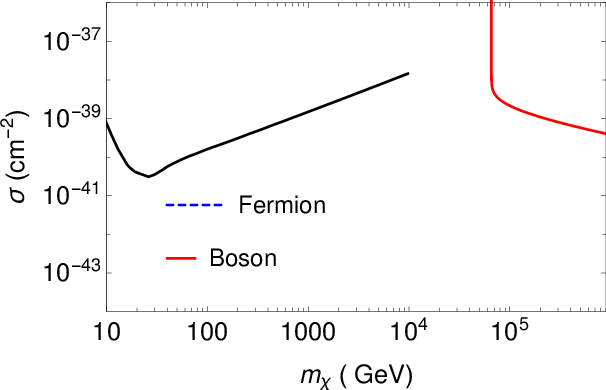}}
\hspace{0.05\textwidth}
\subfigure[Cored DM profile (eq. \ref{core}): $\rho =410$ GeV.cm$^{-3}$.]{\includegraphics[width=2.8in,angle=0]{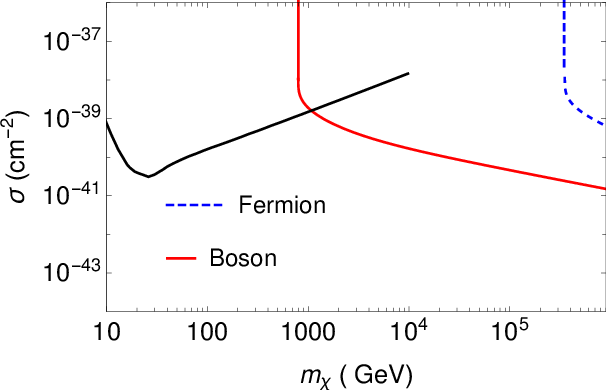}}
\caption{\label{pspace}Constraints on the DM-nucleon scattering cross section $\s$, as a function of the PDM particle mass $m_{\chi}$ for the possible transmutation. The parameter space located on the right side of the red/blue dashed curve suggests a transmutation for $t_{\rm d} \leq t_U$ (implying a transmutation has possibly happened). On the other hand, the left side of the red/blue dashed curve suggests that the transmutation may occur in the future. The black line (equivalent to the black line of Fig. 8 of \cite{aal}) are due to the constraints given by the LZ experiment for the nominal upper limit on the DM-proton spin-dependent cross section as a function of
$m_{\chi}$. Panels (a)-(d) show our results for different DM density profiles.}
\end{center}
\end{figure}

Considering the above and using the general expression of the transmutation timescale $t_{\rm d}$ as discussed earlier, one can calculate $t_{\rm d}$ for WDs.
Fig.~\ref{f5} shows an example of a typical WD in the Galactic bulge ($r \sim 0.1$ pc \cite{bly}), which could transmute into a collapsed object for a wide range of $m_{\chi}$ values, depending on the value of $\rho$. This collapsed object could be a naked singularity for a reasonable value of $T_{\rm WD}$ (see Fig. \ref{fwda}).
More elaborately, a Galactic bulge (with NFW DM profile) WD, having properties like the fastest-spinning ($T_{\rm WD} \approx 25$ s) WD with $M_{\rm WD}=0.7M_{\odot}$, and $R_{\rm WD}=0.01R_{\odot}$ and capturing bosonic (fermionic) PDM particles of $20$ GeV $< m_{\chi} <  10^6$ GeV  ($2\times 10^3$ GeV $< m_{\chi} < 10^6$ GeV) with the value of $\s$ indicated in Panel (a) of Fig.~\ref{pspace}, could actually transmute into a naked-singularity of $a \sim 19$. On the other hand, the same Galactic (but with the Einasto DM density profile) WD could transmute into a naked-singularity of the same spin ($a \sim 19$) by capturing bosonic (fermionic) PDM particles of $60$ GeV $< m_{\chi} <  10^6$ GeV  ($ 10^4$ GeV $< m_{\chi} < 10^6$ GeV) with the value of $\s$ indicated in Panel (b) of Fig.~\ref{pspace}. This indicates that our result varies slightly (see also Figs. \ref{f5} and \ref{pspace}) for considering the NFW and Einasto DM density profiles. 
However, if the same WD is located in the Galactic disk (i.e., DM density $\rho \sim 0.4$ GeV.cm$^{-3}$ \cite{hoo}), we get $t_{\rm d} > t_{\rm U}$ for almost all values of $m_{\chi}$.
Thus, we might not find either a naked-singularity or a BH, which is transmuted from a WD, in the Galactic disk. 
Similarly, one may not find naked singularities and BHs transmuted from WDs even in the Galactic bulge for $\s < 10^{-43}$ cm$^{2}$ (see Fig.~\ref{pspace}), because the curves in Fig.~\ref{f5} would be above the $t_{\rm d} = t_{\rm U}$ line. For the cored DM density profile with $r_{\rm c}=1$ kpc, Panel (c) of Fig.~\ref{f5} shows that the transmutation could not be possible in the reasonable time, if the PDM particles are fermions. In case of the bosonic PDM particles, the transmutation could be possible for the mass range: $6 \times 10^4$ GeV $< m_{\chi} <  10^6$ GeV with the value of $\s$ indicated in Panel (c) of Fig.~\ref{pspace}. Panel (d) of Fig.~\ref{f5} with the cored DM density profile  of $r_{\rm c}=0.2$ kpc shows that the same WD could trnasmutre to a naked singularity by capturing bosonic (fermionic) PDM particles of $730$ GeV $< m_{\chi} <  10^6$ GeV  ($3\times 10^5$ GeV $< m_{\chi} < 10^6$ GeV) with the value of $\s$ indicated in Panel (d) of Fig.~\ref{pspace}.
Comparing Panels (a)-(d) of Fig.~\ref{f5} one can conclude that the curves are shifted towards right if $\rho$ decreases. This is also clear from Fig.~\ref{pspace}.

Fig.~\ref{pspace} constrains the $\s - m_{\chi}$ parameter space for the possible transmutation within the present age of the Universe, and Panels (a)--(d) show  this $\s - m_{\chi}$ parameter space for different DM density profiles. The parameter space located on the right side of the red (or, blue dashed) curve is for  $t_{\rm d} \leq t_U$ (which suggests that the transmutation has already occurred).
On the other hand, the left side of the red (or, blue dashed) curve suggests that the transmutation may occur in the future, as $t_{\rm d} > t_U$. 
 In Panel (c) of Fig.~\ref{pspace}, we consider a cored DM density profile with core radius $1$ kpc \cite{oll}. The blue dashed curve (for fermion) of the same is outside the parameter space shown in the figure. This indicates that the transmutation could not be possible for this specific case, if the PDM particles are fermions. On the other hand, the cored DM density profile with core radius $0.2$ kpc \cite{oll} shows that the solid red and blue dashed curves are shifted towards right compared to the similar curves of Panels (a) and (b) of Fig.~\ref{pspace}. However, the consideration of different DM density profiles does not change our conclusion. 

The black line of Fig.~\ref{pspace} (equivalent to black line of Fig. 8 of \cite{aal}) are from the data provided in the Supplemental Material of \cite{aal}. This black line is due to the constraints given by the LUX-ZEPLIN (LZ 2022) experiment for the nominal upper limit on the DM-proton spin-dependent cross section as a function of $m_{\chi}$. 
Note that we do not specifically consider here a spin-independent or spin-dependent DM-nucleon interaction. However, the possibility of the transmutation of WDs into BHs/naked singularities is more for spin-dependent DM-nucleon interaction. 
Thus, in Fig.~\ref{pspace}, we show the constraints for the spin-dependent case.
Such spin-dependent interaction could happen due to nuclei with nonzero spin, for example, C$^{13}$ \cite{ktwd}.
It is important to note here that the overall mass range of the PDM particles in the LZ \cite{aal} and XENONnT \cite{apr} DM search experiments is considered upto $m_{\chi} \sim 10^4$ GeV (as indicated by the black line in Fig.~\ref{pspace} as well) and a standard halo model with parameter values fixed (e.g., $\rho=0.3$ GeV.cm$^{-3}$) to the recommendations of \cite{bax} is assumed. 
Thus, these experimental results do not constrain our results above $10^4$ GeV 
(in this paper, we consider the PDM particles' mass upto $m_{\chi} \sim 10^6$ GeV), 
and 
the conclusions of these experiments depend on the assumed values. 
Our results show that there are ample possibilities of transmutation of WDs into BHs/naked singularities.
Moreover, the consideration of different DM density profiles does not change our
conclusion.

\section{Conclusion and Discussions}
In conclusion, it is evident that many WDs could realistically transmute into Kerr naked singularities or BHs, and the uncertainties in the WD parameter values do not change our main conclusions. The naked singularities have been theoretically shown to be viable end products of collapse \cite{psj, jmn, ss, jm, gss, jsy, wag, gh}. Some of the collapsed objects could be naked singularities instead of BHs. For example, Sgr A* could be a Joshi-Malafarina-Narayan-1 (JMN-1 \cite{jmn1}) naked singularity \cite{eht2022}, M87* could be a Kerr-Taub-NUT naked singularity \cite{gcyl} or a Kerr naked singularity with $4.5 < a < 6.5$ \cite{bambi, gcyl}, GRO J1655-40 could be a Kerr-Taub-NUT naked singularity \cite{cbgm}.
In a recent paper \cite{pns}, it is also indicated that many of the proposed PBHs could actually be primordial naked singularities. There are some distinguishable effects already predicted \cite{ckp, ckj, jmn} between a BH and a naked singularity. Note that the typical Kerr parameter values of NSs are less than $1$, and hence NSs are likely to transmute into BHs, and not into naked singularities (see Appendix \ref{s3}). 
Moreover, as the mass ejection was shown to be small during the accretion by an endoparasitic BH \cite{el}, a typical NS (mass $\sim 1.4M_\odot$) could transmute into a somewhat super-solar mass BH, and usually not into a sub-solar mass BH.
Thus, the transmutation of an NS could give rise to only a BH within a relatively narrow mass range.
Besides, while the Kerr parameter is greater than $1$ for stars and
planets, the transmutation times for them are typically larger than their
lifetimes and/or $t_{\rm U}$ (see Appendix \ref{s4}). So, it is unlikely that stars and
planets would transmute into either BHs or naked singularities (although some low-mass stars in the Galactic bulge could transmute into naked singularities). 
Therefore, as we show in this paper, WDs are likely the most promising candidates to give rise to near- and sub-solar-mass naked singularities and BHs with a wide mass range via transmutation. 
Note that, while this general conclusion relies on the existence of non-self-annihilating PDM particles and suitable minimum values of the PDM particle-nucleon scattering cross section, it does not depend on an unknown amount of mass ejection during the accretion by the endoparasitic BH, or some other uncertainties related to the complex accretion process, magnetic field, etc.
This is because, while the details of the transmutation would depend on such complex and poorly understood processes and unknown parameter values \cite{mark}, the final products should not be different from those concluded above.  
However, this general conclusion
of the transmutation into a BH or a naked singularity is valid, if the
WD does not undergo a supernova induced by the
accumulated PDM particles. 
But note that, even if there is a
supernova after a tiny BH is formed by the collapse of PDM particles at the core of a WD \cite{jan}, the BH should remain, and there could be a sufficient
amount of fallback matter which might eventually be accreted by the BH
and hence the BH might grow.

Let us now consider the possibility of a supernova induced by the accumulated PDM particles. 
Self-annihilating PDM particles might not allow to accumulate sufficient PDM mass, and even if there is such an accumulation, the energy produced due to annihilation might cause a supernova \cite{raj}. 
But, since we consider non-self-annihilating PDM particles, chances of a supernova is lesser \cite{raj}. 
The possibility of triggering a class of supernova induced by the PBHs and DM was discussed in \cite{gra1, gra2}. 
In particular, the possibility of supernovae for $m_{\chi} > 100$ TeV PDM particles was argued in \cite{jan}.
It was mentioned that the DM, which sufficiently heats a local region in a WD, triggers a runaway fusion process to ignite a type Ia supernova \cite{jan}. 
However, if the above limit on $m_{\chi}$ is valid, the possibility of a supernova does not mostly affect our conclusions because we consider $m_{\chi} \sim 10 $ GeV to a few $100$ TeV range. For example, our Fig. \ref{f5} clearly shows that even if we consider only the $m_{\chi} < 100$ TeV PDM particles, some WDs can be transmuted into BHs or naked singularities. 
Moreover, for non-self-annihilating PDM particles, \cite{ste} showed that an endoparasitic BH is formed inside a WD, and may accrete matter from the WD until a macroscopic BH is formed (although this paper does not consider naked singularities). In support to their claim, \cite{ste} argued that 
the heat could be transferred between the nuclear matter and DM, nuclear burning phases could be stable and carbon could have been depleted early \cite{ste}. 
Support for the formation and growth of a BH can also be found elsewhere (e.g., page 20 of \cite{raj} and references therein; page 5 of \cite{dasg}). 
Thus, in view of the above, we conclude that, while a supernova for our case could be a possibility for high mass PDM particles, the transmutation of a WD into a BH or a naked singularity is a viable consequence of the capture of non-self-annihilating PDM particles by the WD.

The total number of WDs is estimated to be about $10^5$ in a Hubble view of the Galactic bulge \cite{cal}\footnote{https://www.nasa.gov/feature/goddard/hubble-uncovers-the-fading-cinders-of-some-of-our-galaxy-s-earliest-homesteaders}, which is more than the expected number of NSs \cite{kd}.
This makes the new channel to form near- or sub-solar-mass naked singularities and BHs viable, and indicates that several such objects could exist in the Galactic bulge.
In fact, it has recently been shown that there are infinitely many boundary conditions that make the Kerr superspinars/naked-singularities stable \cite{nakao2}, which suggests that such objects could realistically exist. Moreover, the naked singularities do not exhibit similar characteristics to BHs in the strong gravity regime, as evident from \cite{ckp,ckj, cbgm}.

What could be an observational signature of the transmutation of a WD? When a typical WD orbits around and/or is accreted by a BH, an about $10^{49}-10^{52}$ erg energy could be released \cite{fw} that may cause events like ultraluminous X-ray bursts \cite{shen}, Type I supernovae \cite{wm}, and long duration gamma-ray bursts \cite{fw}. In an endoparasitic process, it is possible that a significant amount of matter is not ejected \cite{el}. However, acting as the energy reservoir, the magnetosphere of a WD could release a significant amount of energy in $\sim 1$ ms during the collapse. For example, for a highly magnetic WD of surface magnetic field $B_{\rm WD} \sim 10^9$ G \cite{dm}, a short burst of luminosity \cite{leh} $L \sim B_{\rm WD}^2 R_{\rm WD}^3/$ms $\approx 10^{47}$ erg/s could be radiated to infinity as an observable transient signal \cite{el}.

The current gravitational wave observatories, as well as the future ones (aLIGO \cite{aligo}, LISA \cite{lisa}), provide a promising way to detect near- and sub-solar-mass naked singularities and BHs.
How can one test or falsify that observed near- and sub-solar-mass BHs are transmuted WDs and not PBHs?
In case of the former, the merger rates involving such collapsed objects over the cosmological time scale (sufficient data of which may be available in the future) should be correlated with the formation of stars of relevant masses over the same time scale \cite{madau} and such mergers would be found only from galactic bulges or similar dense regions.
Such conditions may not hold for the latter option.

Moreover, transmutation should give rise to low mass BHs with a range of $a$ (Kerr parameter) values, some with very high $a$. On the other hand, we probably expect relatively low $a$ values for PBHs  \cite{luca, mir}. This could be another way to observationally distinguish transmuted WDs from PBHs.

\appendix

 \begin{center}
  {\bf APPENDIX}
 \end{center}

 \section{\label{s3} Can neutron stars transmute into Kerr naked singularities or Kerr black holes?}
\begin{figure}
 \begin{center}
\subfigure[$\o=10^3$ Hz]{\includegraphics[width=2.6in,angle=0]
{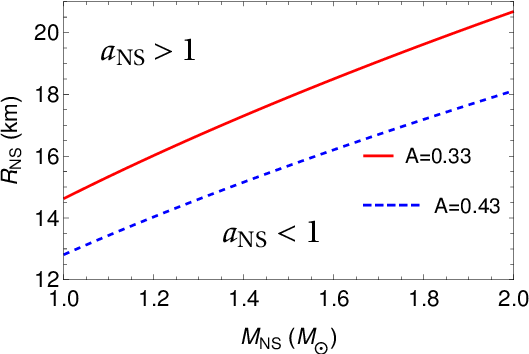}}
\hspace{0.05\textwidth}
\subfigure[$\o=700$ Hz]
{\includegraphics[width=2.6in,angle=0]{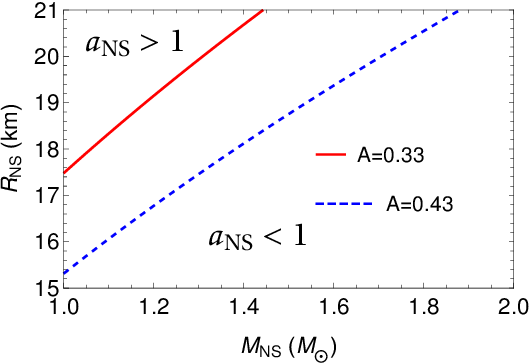}}
\caption{\label{f1}
Dependence of the Kerr parameter ($a_{\rm NS}$) on NS radius and mass, for a given spin frequency (see Eq. \ref{feq}).
The solid red and dashed blue curves represent $a_{\rm NS}=1$ for $A=0.33$ and $A=0.43$, respectively.}
\end{center}
\end{figure}

Using Eq. (\ref{feq}) for neutron stars (NSs), we show the parameter spaces corresponding to BH (NS Kerr parameter $a_{\rm NS} \leq 1$) and naked singularity ($a_{\rm NS} > 1$) in Fig. \ref{f1}. We use $A = 0.33$ and $0.43$ \cite{bha17}.
Although the spin frequency ($\o$) of the fastest known pulsar is $716$ Hz \cite{hes}, the break-up spin frequency is typically around $1250-1500$ Hz \cite{bha16} depending on the stellar equation of state. 
Therefore, we plot our results for $\o=10^3$ Hz and $\o=700$ Hz in 
Panels (a) and (b) of Fig. \ref{f1}, respectively.
Considering the realistic NS parameter values \cite{bha16, miller, ozel}, it appears from Fig. \ref{f1} that NSs have $a_{\rm NS} < 1$, and hence they cannot transmute into naked singularities.

\begin{figure}
 \begin{center}
\subfigure[$\s \geq 10^{-44}$ cm$^2$]{\includegraphics[width=2.6in,angle=0]
{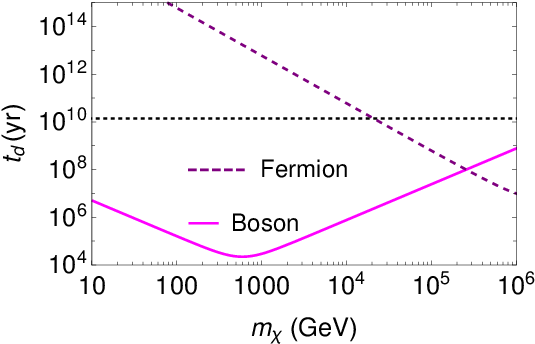}}
\hspace{0.05\textwidth}
\subfigure[$\s=10^{-52}$ cm$^{2}$]
{\includegraphics[width=2.6in,angle=0]{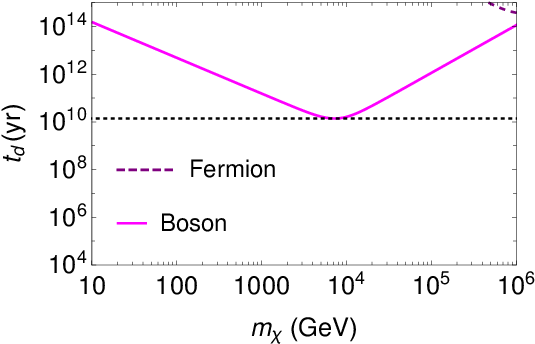}}
\caption{\label{f2}
Similar to Fig. \ref{f5}(a), but for an NS of $M_{\rm NS}=1.6M_{\odot}$ and $R_{\rm NS}=16.5$ km in the Galactic bulge with the NFW DM density profile. }
\end{center}
\end{figure}

\begin{figure}
 \begin{center}
{\includegraphics[width=2.6in,angle=0]{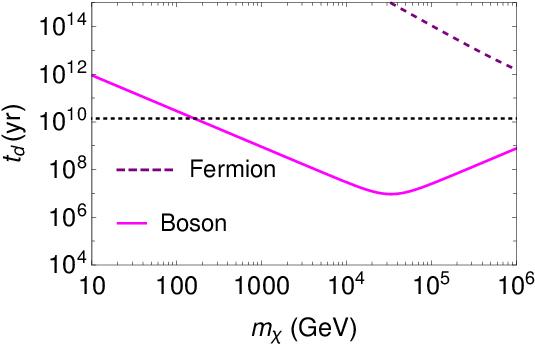}}
\caption{\label{f3}Similar to Fig. \ref{f2}(a) with $\s \geq 10^{-44}$ cm$^2$, but the NS is located in the Galactic disk.}
\end{center}
\end{figure}

Can NSs realistically transmute into BHs? To calculate the transmutation time $t_{\rm d}$ for a spinning NS, we follow the same procedure as for a WD. We summarize it below. In case of an NS, one obtains \cite{mcd}
\begin{eqnarray}
 N_{\rm self}^{\rm NS}=4.8 \times 10^{41} \left(\f{100~{\rm GeV}}{m_{\chi}}\right)^{5/2}\left(\f{\T}{10^5~{\rm K}}\right)^{3/2},
\end{eqnarray}
where $\T \sim 10^5$ K \cite{mcd}.
Note that we do not consider the Bose-Einstein condensate (BEC) formation, as the required $\T$ is always less than the core temperature of a stellar object \cite{mcd, dasg}. With this information, one can  calculate the formation timescale $t_0^{\rm b}=t_{\rm th}+N_{\rm self}^{\rm NS}/F$ (or, $t_0^{\rm f}=t_{\rm th}+N_{\rm Ch}^{\rm f}/F$) of an endoparasitic BH inside an NS due to the accumulation of bosonic (or, fermionic) PDM particles, where $t_{\rm th}$ for a NS can be estimated using Eq. (20) of \cite{mcd}. We plot $t_{\rm d}$ for various values of $m_{\chi}$ in Fig. \ref{f2} and Fig. \ref{f3} by considering the NFW DM density profile. These figures show that an NS could realistically transmute into a collapsed object (i.e., a BH).

\section{\label{s4}Can other cosmic objects transmute into Kerr naked singularities or Kerr black holes?}
The lifetimes of high-mass stars are only about millions of years.
Thus, they do not get enough time to form endoparasitic BHs.
If an earth-like planet (sun-like star) transmutes into a collapsed object without a significant decrease of the Kerr parameter $a$, this object would be a Kerr naked singularity with $a \approx 890~ (1.2)$. 
However, for an earth-like planet and a sun-like star in the Galactic disk, the transmutation time would be at least $\sim 10^{16}$ yr, i.e., $t_{\rm d} >> t_{\rm U}$. 
An earth-like planet cannot be transmuted into a naked singularity in a reasonable time even if it is located in the Galactic bulge as $t_{\rm d}$ is at least $\sim 10^{15}$ yr. However, if a sun-like star is located in the Galactic bulge, 
it could transmute into a naked singularity in $t_{\rm d} \sim 10^{11}$ yr. 
\\

{\bf Acknowledgements :}  We thank the referee for constructive comments that helped to improve the manuscript. One of us (CC) thanks Anupam Ray for some useful discussions. CC dedicates this paper to his mother Shree Snigdha Chakraborty who was his constant inspiration in all fields of life, and unfortunately passed away on September 11, 2023.

\end{document}